\documentclass[12pt,preprint]{aastex}

\shorttitle{Understanding Lyman-$\alpha$ Nebulae at Low Redshift I}

\newcommand{\lya}{Ly$\alpha$}
\newcommand{\neiii}{[\hbox{{\rm Ne}\kern 0.1em{\sc iii}}]}
\newcommand{\neiv}{[\hbox{{\rm Ne}\kern 0.1em{\sc iv}}]}
\newcommand{\nev}{[\hbox{{\rm Ne}\kern 0.1em{\sc v}}]}
\newcommand{\sii}{[\hbox{{\rm S}\kern 0.1em{\sc ii}}]}
\newcommand{\nii}{[\hbox{{\rm N}\kern 0.1em{\sc ii}}]}
\newcommand{\nv}{\hbox{{\rm N}\kern 0.1em{\sc v}}}
\newcommand{\oi}{[\hbox{{\rm O}\kern 0.1em{\sc i}}]}
\newcommand{\oii}{[\hbox{{\rm O}\kern 0.1em{\sc ii}}]}
\newcommand{\oiii}{[\hbox{{\rm O}\kern 0.1em{\sc iii}}]}
\newcommand{\halpha}{\hbox{{\rm H}\kern 0.1em$\alpha$}}
\newcommand{\hbeta}{\hbox{{\rm H}\kern 0.1em$\beta$}}
\newcommand{\hgamma}{\hbox{{\rm H}\kern 0.1em$\gamma$}}
\newcommand{\hdelta}{\hbox{{\rm H}\kern 0.1em$\delta$}}
\newcommand{\hei}{\hbox{{\rm He}\kern 0.1em{\sc i}}}
\newcommand{\heii}{\hbox{{\rm He}\kern 0.1em{\sc ii}}}

\newcommand{\numhighbeans}{9}
\newcommand{\numlowbeans}{7}
\newcommand{\GBfour}{J0113$+$0106}

\newcommand{\GBsix}{J1155$-$0147}

\newcommand{\GBsixteen}{J2240$-$0927} 
\usepackage{color}
\usepackage{adjustbox}

\begin{document}

\title{Understanding Lyman-$\alpha$ Nebulae at Low Redshift I: The Sizes, Powering, and Kinematics of ``Green Bean'' Galaxies}

\author{Moire K. M. Prescott\altaffilmark{1} \& Kelly N. Sanderson\altaffilmark{1}}
\altaffiltext{1}{Department of Astronomy, New Mexico State University, P. O. Box 30001, MSC 4500, Las Cruces, NM, 88003, USA; mkpresco@nmsu.edu}

\begin{abstract}

A new but rare sample of spatially extended emission line nebulae, nicknamed ``Green Beans'', 
was discovered at $z\approx0.3$ thanks to strong \oiii\ emission, 
and subsequently shown to be local cousins of the Lyman-$\alpha$ (\lya) nebulae found at high redshift.  
Here we use follow-up APO/DIS spectroscopy to better understand how these low redshift \lya\ nebulae 
compare to other populations of strong emission line sources.  
Our spectroscopic data show that low-z \lya\ nebulae have AGN-like emission line ratios, 
relatively narrow line widths (${\rm FWHM}\lesssim1000$ km s$^{-1}$), and emission line kinematics 
resembling those of Type~2 AGN at the same redshift, confirming that they 
are powered by Type~2 AGN with typical ionizing continua.  
While low-z \lya\ nebulae are larger and less concentrated than compact, star-forming 
Green Pea galaxies, we find that they resemble typical Type~2 AGN in terms of $r$-band 
concentration and size.  
Based on this pilot study, low-z \lya\ nebulae appear to be a subset of Type 2 AGN 
with bluer optical continua and high \oiii\ equivalent widths but with 
comparable sizes and similar \oiii\ kinematics.  
These characteristics may simply reflect 
the fact that low-z \lya\ nebulae are drawn from the high luminosity end of the Type 2 AGN 
distribution, with higher nuclear activity driving higher \oiii\ equivalent widths 
and more central star formation leading to bluer optical continua.  
Deeper spectroscopic follow-up of the full sample will shed further light on these issues and on 
the relationship between these low-z \lya\ nebulae and the \lya\ nebula population at high redshift.

\end{abstract}


\section{Introduction}
\label{sec:intro}

The circumgalactic medium lies at the interface between galaxies and the surrounding intergalactic medium.  
In cases where the circumgalactic medium is lit up as extended emission line 
nebulae, we gain spatially resolved information about both the gas reservoir 
available to fuel star formation and the effects of feedback from the central galaxy.  
\citet{schirmer2013} discovered a new sample of extended emission line nebulae in the local universe 
using photometry from the Sloan Digital Sky Survey \citep[SDSS;][]{york2000}.  
Due to the presence of strong \oiii\ line emission in the $r$-band, these objects showed up as outliers 
in $gri$ color space (and green in composite SDSS $gri$ imaging), a region that is otherwise dominated by image artifacts.  
In contrast to the more well-known, compact ``Green Pea'' galaxies \citep{cardamone2009}, these objects were 
selected to be spatially extended by at least 2\arcsec.  Dubbed ``Green Beans,'' therefore, 
due to their larger sizes, 17 of these 
objects were found at $0.19\lesssim z \lesssim0.35$.  
Their high \oiii/\hbeta\ ratios combined with a lack of any detected broad emission 
lines suggested powering by Type 2 Active Galactic Nuclei (AGN).  Subsequent analysis of 
Galaxy Evolution Explorer (GALEX) data revealed that, in addition to showing spatially extended \oiii\ 
emission, these Grean Beans also emit strong ultraviolet emission consistent with being \lya\ 
at these redshifts \citep{schirmer2016}.  
Recent Hubble Space Telescope imaging and slitless spectroscopy has confirmed the presence of 
\lya\ emission \citep{keel2019}, providing strong support for the idea that Green Beans 
are in fact lower redshift cousins of the extended \lya\ nebulae seen at high redshift ($z\sim2-6$).  

With a space density of $\sim$4.4 Gpc$^{-3}$, these low redshift \lya\ nebulae 
are extremely rare, even more so than \lya\ nebulae at high redshift \citep{schirmer2013}.  
This fact raises several questions.  In particular, what makes \lya\ nebulae 
such a rare phenomenon at low redshifts, and 
how do they compare to the more well-studied Green Pea galaxies and to the larger population 
of Type 2 AGN at similar redshifts? 
Furthermore, what does the declining space density of these 
extended emission line regions from high to low redshift 
imply about the evolution of the circumgalactic medium or the physics 
of galaxy formation over cosmic time?  
To begin to address these questions, we are obtaining follow-up 
low ($R\sim600-1500$) and medium ($R\sim2500-4500$) 
resolution spectroscopy of low-z \lya\ nebulae 
with the Apache Point Observatory (APO) 3.5m telescope and the 
Dual Imaging Spectrograph (DIS).  

This paper presents results from our pilot program targeting a 
subset of the $z\approx0.3$ \lya\ nebula sample.  
We describe the pilot sample in Section~\ref{sec:sample} and our observations 
and reductions in Section~\ref{sec:data}.  
Section~\ref{sec:fitting} details our emission line fitting analysis, and 
in Section~\ref{sec:results}, we investigate the powering mechanism, colors, 
sizes, concentrations, and kinematics of 
these low-z \lya\ nebulae.  In Section~\ref{sec:discussion} we discuss 
how low-z \lya\ nebulae relate to other emission line populations at similar redshifts, 
in particular, star-forming Green Pea galaxies and typical Type 2 AGN, as well as what we can learn from comparisons 
to extended \lya\ nebulae at higher redshifts. 
We conclude in Section~\ref{sec:conclusions}.  
Throughout, we assume the standard $\Lambda$CDM cosmology ($\Omega_{M}$=0.3, $\Omega_{\Lambda}$=0.7, $h$=0.7);
the angular scale at $z\approx0.3$ is 4.494~kpc arcsec$^{-1}$.  
We adopt vacuum wavelengths for the analysis, following the SDSS approach, 
but we quote line names using standard air wavelength designations, e.g., \oiii$\lambda$5007.  
All magnitudes are in the AB system \citep{oke74}.


\section{The Target Sample}
\label{sec:sample}

Out of the 17 \lya\ nebulae reported to be at $z\approx0.3$ in \citet{schirmer2013}, 
we were able to target \numhighbeans\ (\numlowbeans) 
at medium (low) spectral resolution during our initial pilot program.  
Standard IAU designations for the target objects are listed in Table~\ref{tab:obs}; 
we use an abbreviated form (e.g., \GBfour) throughout the rest of the paper.
Based on follow-up observations by \citet{schirmer2016}, the objects in the pilot sample show 
a range of environments -- some appear isolated while others reside within galaxy groups -- 
and have \oiii\ nebula diameters of 25-75~kpc.  
The X-ray fluxes of the observed 
low-z \lya\ nebulae range from the most luminous X-ray source in the 
sample to one of the least luminous, and more than half show hints of outflow and/or merger 
activity based on ground-based optical imaging.  
Two of the objects in our pilot sample (\GBfour, \GBsix) plus one 
additional low-z \lya\ nebula (\GBsixteen) have been targeted with 
Gemini/GMOS \citep{davies2015,kawamuro2017}, showing further evidence 
for complex kinematics and outflows. 

\section{Data}
\label{sec:data}

\subsection{Spectroscopic Observations}
\label{sec:specobs}

Spectroscopic observations of the target low-z \lya\ nebula 
sample were obtained using the Apache Point Observatory 3.5m telescope and 
the DIS spectrograph on UT 2018 January 12 and 24, April 15 and 22, and May 9.  
Position angles were chosen to cover the largest extent of the nebula, based 
on existing deep optical imaging \citep{schirmer2016}, and to include nearby sources on the slit.
The DIS native pixel scale is 0.42\arcsec/pixel and 0.40\arcsec/pixel for the 
blue and red sides, respectively.  
During these observations, DIS was affected by a scattered light issue caused by 
condensation on the dewar window; we discuss this further in Section~\ref{sec:specredux}.  
Details on the observations are listed in Table~\ref{tab:obs}.  

For the first four nights (UT 2018 January 12 and 24, April 15 and 22), 
we used the medium resolution B1200 and R1200 gratings and a 1.5\arcsec\ slit, 
yielding a resolution element of $\sim1.5-2.0$\AA\ and a resolving power of $R\sim2380-4650$.  
The grating central wavelengths were set to 4800\AA\ and 6600\AA\ for the blue and red sides, 
respectively, 
on UT 2018 January 12, January 24, and April 15, while on UT 2018 April 22, the blue and red grating central 
wavelengths were shifted to 4562\AA\ and 6137\AA, respectively, in order to target two of the lower 
redshift \lya\ nebula systems. 
For each object, we obtained 2-6 exposures of 300-600s each, with occasional dither offsets of $\approx5$\arcsec\ in 
between individual exposures, and took data for two different position angles for each object, when possible.  
Conditions on UT 2018 January 12 and 24 were clear, with seeing of 1.0-1.6\arcsec\ and 
1.5-1.9\arcsec, respectively. 
On UT 2018 April 15 conditions were clear with seeing of 1\arcsec, but 
the data were inadvertently binned spatially by 2, leading to a pixel scale of 
0.84\arcsec/pixel and 0.80\arcsec/pixel on the blue and red sides, respectively.  
Conditions were non-photometric on UT 2018 April 22, with seeing of 1.3\arcsec\ and intermittent clouds.   

For the final night on UT 2018 May 9, we used the lower resolution B400 and R300 gratings and a 1.5\arcsec\ slit, 
yielding a resolution element of $\sim5.7-6.5$\AA\ and a resolving power of $R\sim630-1425$.  
The blue and red grating central wavelengths were set to 
their nominal values of 4400\AA\ and 7500\AA, respectively, providing full coverage of the optical window.  
We observed only one position angle for each object, choosing the one covering the larger spatial dimension of the 
nebula, with 3-4 exposures of 200-300s each.  The conditions were clear with a native seeing of 
1.2-1.3\arcsec, but quickly falling temperatures led to unstable focus during the first part of the night.

\subsection{Spectroscopic Data Reduction}
\label{sec:specredux}

We carried out the data reduction in the usual manner with {\it IRAF}.  
We applied bias and flat-field calibrations, and then shifted and combined multiple exposures.  
Wavelength calibration was applied using HeNeAr calibration lamp exposures 
taken at the position of the target, and checked against sky lines,  
yielding a wavelength solution with an rms accuracy of 0.07\AA\ (0.7\AA) 
for the medium (low) resolution spectra. 

We flux-calibrated the data using observations of the spectrophotometric standard stars G191B2B 
(UT 2018 January 12), Feige~34 (UT 2018 January 24), BD+3332642 (UT 2018 April 15), and 
HZ44 (UT 2018 May 9), along with the tabulated atmospheric extinction from 
APO\footnote{http://astronomy.nmsu.edu:8000/apo-wiki}.  
The observations from UT 2018 April 22 were reduced along with the rest of the dataset, but could not 
be flux-calibrated directly due to non-photometric conditions.  These data were used to derive 
measurements for which absolute flux calibration is not required (i.e., line FWHMs and equivalent 
widths, line ratios, and non-parametric \oiii\ kinematic parameters).  

Using the 2D reduced spectra, we extracted a 1D spectrum for each object using 
an aperture width of 8 native, unbinned pixels (3.2\arcsec), 
chosen to approximately match the SDSS-III 
Baryon Oscillation Spectroscopic Survey (BOSS) 3\arcsec\ diameter fibers, 
and centered on the spatial centroid of the \oii$\lambda$3727 and 
\oiii$\lambda$5007 lines for the blue and red sides, respectively.  
The medium resolution spectra show lines of \nev$\lambda\lambda$3346,3426, 
\neiii$\lambda\lambda$3869,3968, \hei$\lambda$3888,
\oii$\lambda$3727 (blended doublet), 
\hdelta$\lambda$4102, \hbeta$\lambda$4861, \oiii$\lambda\lambda$4959,5007, and 
\heii$\lambda$4685.  
The lower resolution data provide additional detections of \hgamma$\lambda$4340, 
\oiii$\lambda$4363, \oi$\lambda$6300, \halpha$\lambda$6563, \nii$\lambda$6548,6583, 
and, in several of the targets, \sii$\lambda$6716,6731.  

In this paper, we focus on using line ratios and emission line profile measurements 
that are less sensitive to slit losses or scattered light effects.  
We note that slit losses and the scattered light issue do not affect line 
ratios measured from the same side of the DIS instrument.  However, since the scattered 
light effect was greater for the blue versus the red side data, 
it introduces a small offset in the ratios of lines observed on the blue versus 
the red sides of DIS.  Relevant for this paper, we apply a correction to the 
low resolution \nev$\lambda$3426/\oiii$\lambda$5007 line ratio measurements presented 
in Section~\ref{sec:powering}, but note that 
this correction (0.015 dex) is less than the mean line ratio measurement 
error for the pilot \lya\ nebula sample (0.090 dex). 

To enable a consistent comparison between the observed properties of the samples of 
emission line objects considered in this work, we corrected all spectra 
for Galactic extinction using $E(B-V)$ values 
from \citet{schlafly2011}, the standard extinction curve 
from \citet{fitzpatrick1999}, and $R_{V}$=3.1.  
The typical Galactic extinction is $E(B-V)$=0.038$\pm$0.022 for 
the full low-z \lya\ nebula sample. 

\subsection{Supplementary Data}
\label{sec:sdssdata}

Existing photometric measurements -- SDSS broadband model magnitudes (corrected for Galactic extinction), 
$r$-band Petrosian radii, and $r$-band concentration ($r_{90}/r_{50}$) -- 
were taken from the Sloan Digital Sky Survey DR15 \citep{aguado2019} for 
the low-z \lya\ nebula sample as well as for two comparison 
samples at similar redshifts: compact Green Pea galaxies \citep{cardamone2009} and Type 2 AGN \citep{yuan2016}.  
The Green Pea galaxies were selected in a manner similar to the low-z \lya\ nebula 
(``Green Bean'') sample, i.e., to have green colors in the SDSS $gri$ 
imaging due to strong \oiii\ emission within the $r$-band, but with compact sizes ($r_{petrosian}<$2\arcsec).  
Here, we use the Green Pea sample that was classified as 
star-forming in the discovery paper \citep{cardamone2009}.  
The Type 2 AGN sample was selected from 
the SDSS-III/BOSS spectroscopic database by requiring \oiii\ restframe equivalent widths of 
$>$100\AA\ and AGN-like line ratios in the standard \oiii/\hbeta\ vs. \nii/\halpha\ 
diagnostic diagram \citep{yuan2016}. 
To allow for a consistent comparison, we restrict our analysis 
throughout the paper to sources at the same redshift range as 
the low-z \lya\ nebula sample ($0.19<z\le0.35$). 

To supplement the new spectroscopic data from APO/DIS, 
we also downloaded a low resolution ($R\sim1800$) spectrum of one target, \GBfour, available 
from the SDSS Legacy archive \citep{york2000}.  We note that \GBfour\ was the only low-z \lya\ nebula 
that was spectroscopically targeted by SDSS; despite similar properties, \GBfour\ was not 
included in the Type 2 AGN sample because the latter was drawn from BOSS rather than SDSS 
Legacy spectroscopy \citep{yuan2016}.  Finally, we downloaded the existing spectra from SDSS/BOSS 
for the star-forming Green Pea and Type 2 AGN samples.

Emission line flux measurements for the comparison samples were derived 
from the SDSS/BOSS pipeline \citep{bolton2012} for the Type 2 AGN sample and from the 
SDSS Portsmouth emission line catalog \citep{thomas2013} for the Green Peas.  
We applied a correction for Galactic extinction to the Type 2 AGN emission line measurements 
in the same manner as above, 
with a typical Galactic extinction of $E(B-V)$=0.024$\pm$0.018.  
Since the catalog Green Pea emission line measurements already include a correction for 
both internal extinction and Milky Way foreground, we recomputed the emission line fluxes 
including the correction for Galactic extinction only.  
The typical Galactic extinction was $E(B-V)$=0.026$\pm$0.017 
for the Green Pea sample.  
Finally, we also collected emission line flux measurements, corrected for Galactic extinction, 
for \GBsixteen\ that were published previously \citep{schirmer2016}.  
 
\section{Fitting the Emission Line Profiles}
\label{sec:fitting}

To derive flux estimates for all emission lines and allow for direct comparisons to the other emission line samples, 
we first employ a simple Gaussian fitting approach, as was done for the SDSS Portsmouth emission line 
catalog and the SDSS/BOSS spectroscopic pipelines.  
We use the lower ionization \oii$\lambda$3727,3729 doublet, which is only marginally resolved (unresolved) 
in our medium (low) resolution spectra, to constrain the systemic velocity \citep{zakamska2003,boroson2005}. 
 
Lines that are well separated are each fit with a single Gaussian profile.  
In the low resolution data, the \neiii$\lambda$3869+\hei, \hgamma+\oiii$\lambda$4363, 
\halpha+\nii$\lambda\lambda$6548,6584, and \sii$\lambda$6717,6731 line groups are 
each fit simultaneously, with the relative wavelength offsets 
constrained to match the laboratory values, corrected to the systemic redshift.  
We require each of the individual lines in the \nii\ and \sii\ doublets to have the same width, and 
we further constrain the \nii$\lambda\lambda$6548,6584 doublet lines to the theoretical flux ratio 
of 3.0 \citep{storey2000}, consistent with the approach taken by the BOSS 
spectroscopic pipeline \citep{bolton2012}. 

To study the kinematics of the narrow-line region gas in more detail, we then use 
the standard approach of 
fitting the observed \oiii\ doublet line profiles with multiple Gaussian components.  
Here, we follow the approach of \citet{yuan2016}, who 
selected the Type 2 AGN sample from BOSS spectroscopy and used the \oiii\ line profiles to derive 
a suite of non-parametric measurements: velocity widths containing 50\%, 80\%, and 90\% of the flux 
($w50 = v75 - v25$, $w80 = v90 - v10$, and $w90 = v95 - v05$), a relative asymmetry 
parameter ($R = \frac{(v95 - v50) - (v50 - v05)}{w90}$; with a positive value indicating a redward skew, 
i.e., a larger red tail), and a dimensionless kurtosis parameter ($r9050=w90/w50$).  
To allow for comparison between the low-z \lya\ nebulae and the Type 2 AGN sample, 
we smoothed our medium resolution data to match the spectral resolution of the BOSS spectra ($R\sim1850$) and 
adapted their published fitting code \citep{yuan2016,zakamska2014,reyes2008} to fit the \oiii\ 
emission lines using up to four Gaussians and derive the non-parametric kinematic measurements. 
In the fitting code, both \oiii\ doublet lines are constrained to have the same fit, 
with a theoretical flux ratio of 2.967; 
an additional Gaussian component is only added if it leads to a decrease in the reduced $\chi^2$ of $\geq10$\%.  

\section{Results}
\label{sec:results}

Using measurements from the literature as well as our new APO/DIS low and medium resolution spectroscopy, 
we investigate the power source, colors, sizes, concentrations, and kinematics of low-z \lya\ 
nebulae as compared to star-forming Green Peas and Type 2 AGN at the same redshifts.

\subsection{Powering by Type 2 AGN}
\label{sec:powering}
\begin{figure}
\includegraphics[angle=0,width=5.5in]{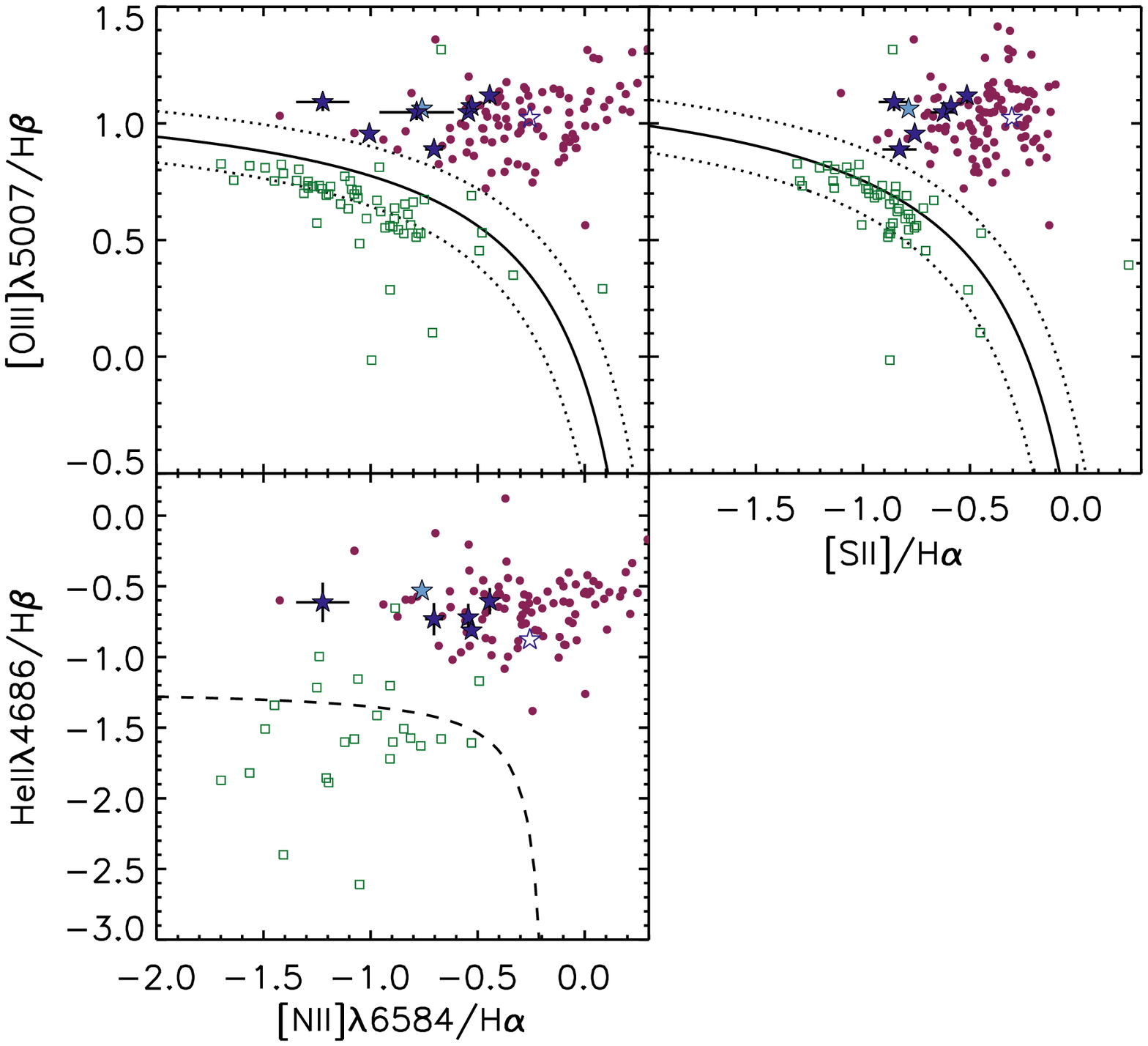}
\caption[]{
Emission line diagnostic diagrams for low-z \lya\ nebulae (blue stars), 
Green Peas \citep[green squares; taken from][]{thomas2013}, and 
Type 2 AGN \citep[magenta circles; taken from][]{bolton2012}. 
(Top) Two of the classic BPT diagrams \citep{bptref} 
with the theoretical maximum star formation line overplotted \citep{kewley2001}.  
The low-z \lya\ nebula 
sample shown includes targets observed in this work (filled dark blue stars) plus one 
object taken from the SDSS archive (\GBfour; filled light blue star), and one object 
(\GBsixteen; open dark blue star) observed by \citet{schirmer2016}. 
Not surprisingly, the low-z \lya\ nebula 
sample lies in the AGN portion of the diagram, overlapping 
the Type 2 AGN sample.  
(Bottom) An alternate AGN vs. star formation classification diagram \citep{shirazi2012}; 
the AGN contribution to the \heii\ emission line increases from lower left to upper right, with 
the dashed line corresponding to a 10\% AGN contribution. 
Again, the low-z \lya\ nebula 
sample lies well into the AGN portion of the diagram, overlapping the 
Type 2 AGN sample.  
Green Peas (green squares) by contrast straddle the 10\% AGN line.  

}
\label{fig:bpt}
\end{figure}

Previous measurements showed that low-z \lya\ nebulae 
have high \oiii/\hbeta\ ratios, consistent with being AGN powered \citep{schirmer2013}.
Using standard BPT diagrams \citep{bptref}, we can further investigate the powering mechanism for a 
subset of the observed low-z \lya\ nebulae 
for which we have low resolution spectra covering 
the entire optical region (Figure~\ref{fig:bpt}).  
As expected, the low-z \lya\ nebulae lie in the upper right of the diagram, 
in the AGN region and above the maximal theoretical star-forming line \citep{kewley2001}.  
For comparison, we include the line ratios for the Type 2 AGN 
and the Green Pea samples.  
While the \oiii/\hbeta\ ratios of low-z \lya\ nebulae are indistinguisable from those of Type 2 AGN, 
the \nii/\halpha\ ratios do appear to be slightly lower than for the Type 2 AGN sample. 
We also plot the alternate selection diagram from \citet{shirazi2012}, which includes \heii\ in place of \oiii, 
and the corresponding dividing line signifying a 10\% contribution from AGN to the \heii\ line; again, 
the low-z \lya\ nebulae are located well inside the AGN portion of the diagram.  
On the other hand, it is clear that despite similarities in terms of how the objects were originally selected 
(and therefore nicknamed), the powering mechanisms for Green Peas and low-z \lya\ nebulae are quite different.  
The star-forming Green Peas show lower \oiii/\hbeta\ and \nii/\halpha\ ratios, 
whereas low-z \lya\ nebulae are AGN powered, with strong \oiii/\hbeta, \nii/\halpha, \sii/\halpha, 
and \heii/\hbeta\ ratios (Figure~\ref{fig:bpt}).

\begin{figure}
\includegraphics[angle=0,width=4.5in]{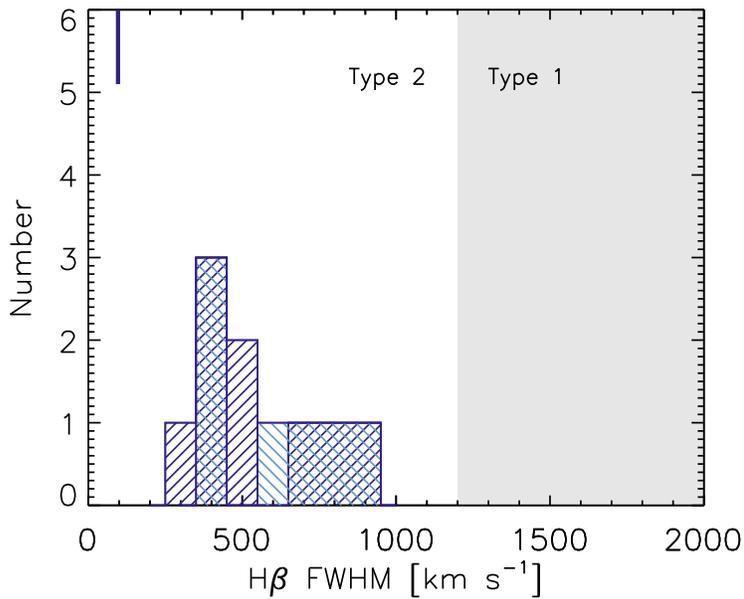}
\caption[]{
Histogram of the \hbeta\ line widths (${\rm FWHM}_{H\beta}$) measured with a single Gaussian 
fit to the medium resolution data and corrected for instrumental resolution.  
Where available, measurements from two different position angles are shown 
(light vs. dark blue hatching), but agree to within the instrumetal resolution in all cases.  
The instrumental resolution of the data is indicated with a blue bar on the top axis. 
A typical dividing line between Type 1 and Type 2 is indicated with grey 
shading \citep[${\rm FWHM}\sim1200$ km s$^{-1}$;][]{hao2003}. 
The low-z \lya\ nebula 
\hbeta\ profiles are spectrally resolved in our data and are relatively narrow, confirming 
that low-z \lya\ nebulae 
are Type 2 objects.
}
\label{fig:allfwhm}
\end{figure}

Previous constraints derived from shallow spectroscopic follow-up of low-z \lya\ nebulae 
could not rule out a broad emission line component to the permitted lines 
\citep[][]{schirmer2013}.  
Using our spectroscopic data, we can measure the typical linewidths for the permitted \hbeta\ 
line in low-z \lya\ nebulae.  
Starting with the single Gaussian fit to the \hbeta\ emission line in our 
medium spectral resolution data and correcting for the instrumental 
resolution, we find a FWHM distribution peaked at $\sim$400 km s$^{-1}$ with 
a tail extending to $\sim$1000 km s$^{-1}$ (Figure~\ref{fig:allfwhm}).  
We check for the presence of a faint broad component 
by refitting the \hbeta\ line using two Gaussian components that are both fixed 
to the central wavelength derived using the single Gaussian fit.  In all cases, 
the hypothetical secondary component has a ${\rm FWHM}<2000$ km s$^{-1}$ (the vast majority are $<1000$ km s$^{-1}$), 
and in no case does the double Gaussian fit provide a substantial ($\geq10$\%) improvement in the reduced $\chi^{2}$. 
Thus, in all targeted low-z \lya\ nebulae, 
we spectrally resolve the \hbeta\ line, do not detect any significant 
broad component, and measure widths well below the typical dividing line between 
Type 1 and 2 AGN \citep[${\rm FWHM}\lesssim1200$ km s$^{-1}$ for permitted lines 
such as \halpha\ or \hbeta; e.g.,][]{zakamska2003,hao2003,mullaney2013}.  
This confirms previous suggestions that low-z \lya\ nebulae 
are Type 2 objects.

\begin{figure}
\includegraphics[angle=0,width=4.5in]{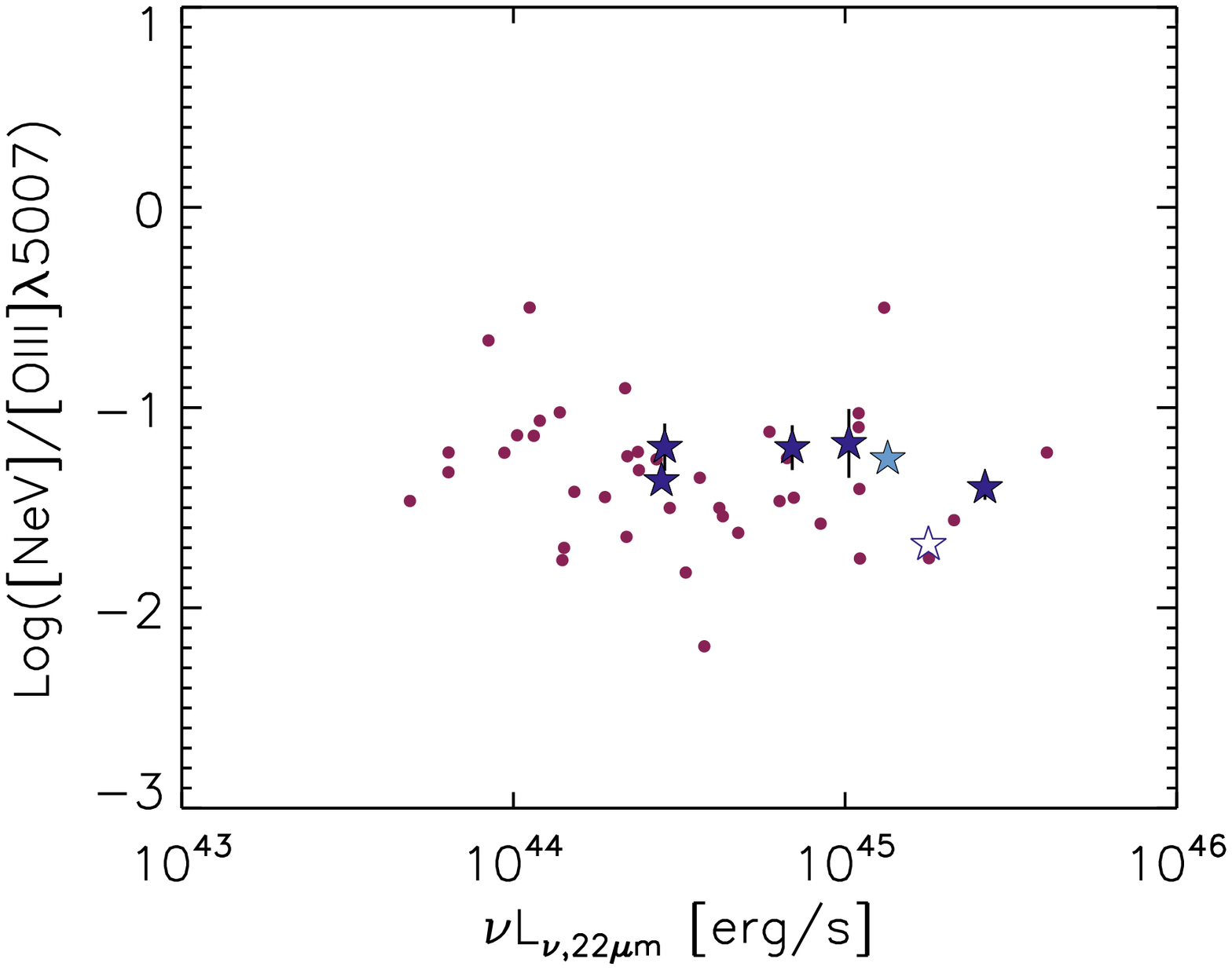}
\caption[]{
WISE 22\micron\ luminosity versus 
\nev$\lambda$3426/\oiii$\lambda$5007 line ratios 
for low-z \lya\ nebulae 
(low resolution data) and Type 2 AGN at the same redshift. 
For objects detected in both emission lines, low-z \lya\ nebulae 
appear to be drawn from the high infrared luminosity end of the distribution but show no significant difference 
in terms of the \nev/\oiii\ emission line ratio, suggesting 
that the ionizing continuum slope of the host AGN is similar in both populations.
}
\label{fig:mirnevoiii}
\end{figure}

Since low-z \lya\ nebulae are powered by Type 2 AGN, we can then ask how the ionizing continuum of 
these objects compares to typical Type 2 AGN.  
Figure~\ref{fig:mirnevoiii} shows the \nev$\lambda$3426/\oiii$\lambda$5007 ratio versus mid-infrared 
luminosities of low-z \lya\ nebulae and Type 2 AGN at similar redshifts.  
While low-z \lya\ nebulae 
appear to have relatively high mid-infrared luminosities, suggesting a powerful 
central engine, we find that their \nev/\oiii\ emission line ratios are quite typical 
when compared to the Type 2 AGN population.  
Since these emission lines both require high but different ionization energies 
\citep[97.2 eV to ionize Ne$^{+3}$ $\rightarrow$ Ne$^{+4}$, 35.1 eV to 
ionize O$^{+}$ $\rightarrow$ O$^{+2}$;][]{NISTASD}, this suggests that the ionizing 
spectral slopes of the host AGN are similar in both populations.

\subsection{Colors and Stacked Spectra}
\label{sec:colors}

\begin{figure}
\includegraphics[angle=0,width=4.5in]{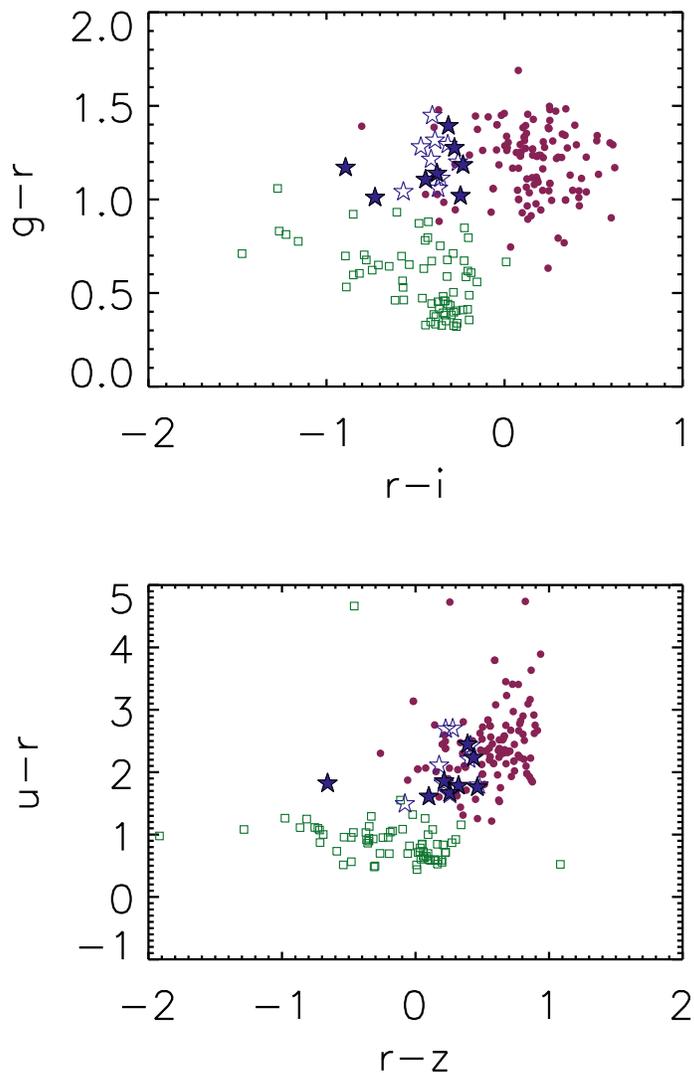}
\caption[]{
Comparison of the broadband colors of Green Pea galaxies (green squares), 
Type 2 AGN (magenta circles), and low-z \lya\ nebulae 
(blue stars), 
all taken from the SDSS archive and corrected for Galactic extinction.  
(Top) $g-r$ versus $r-i$ color-color plot.  
(Bottom)  $u-r$ versus $r-z$ color-color plot.  
All plotted sources are at $0.19<z\le0.35$, a redshift range for which  
the \oiii\ line is contained within the $r$ bandpass.  
The low-z \lya\ nebula 
sample targeted in this paper are indicated as filled blue stars.  

}
\label{fig:type2colors}
\end{figure}

\begin{figure}
\includegraphics[angle=0,width=4.5in]{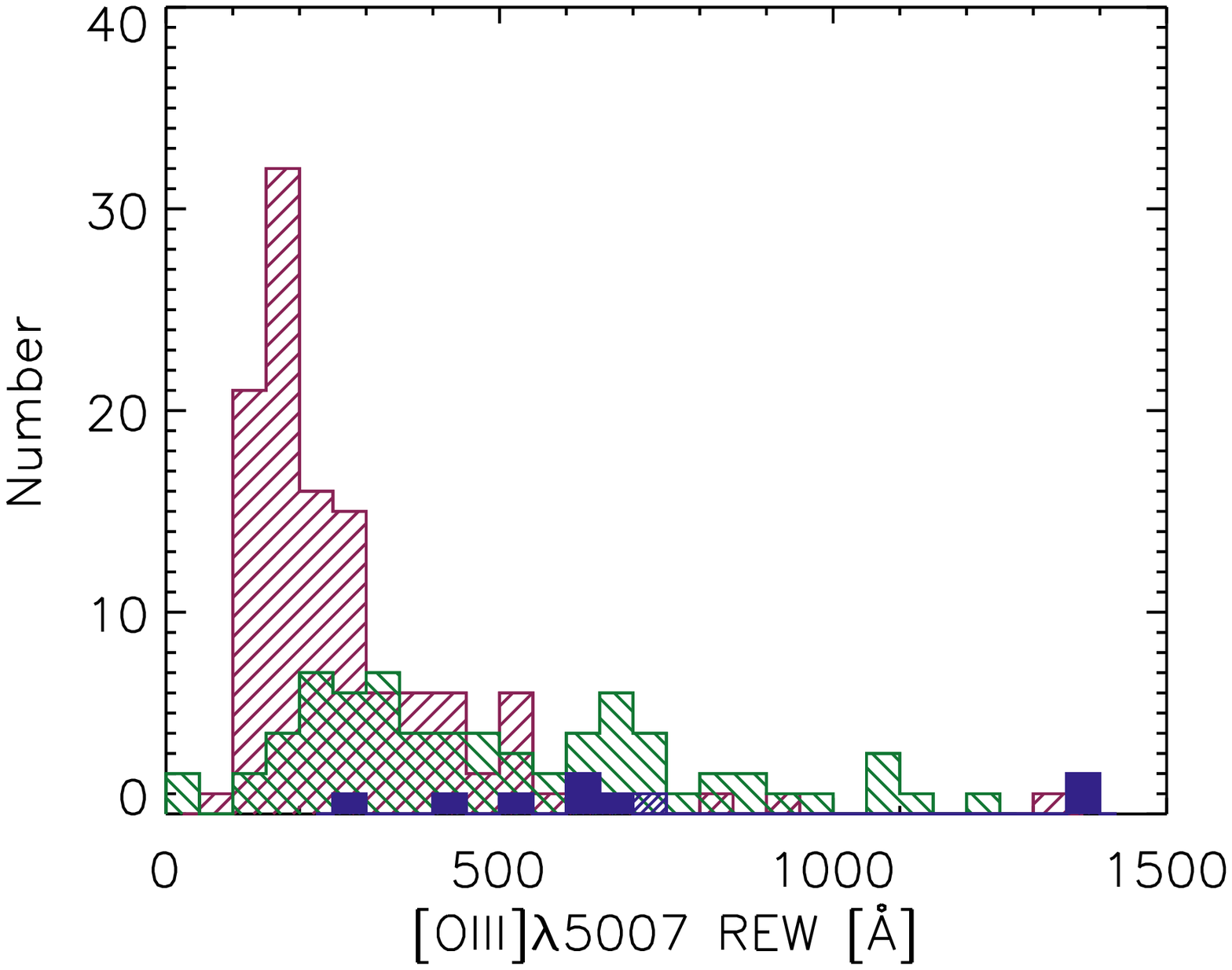}
\caption[]{
\oiii\ restframe equivalent widths reported in the literature for Type 2 AGN 
\citep[hatched magenta histogram;][]{yuan2016} and Green Peas \citep[hatched green;][]{cardamone2009} 
at $0.19<z\le0.35$, as compared to the low-z \lya\ nebula 
pilot sample observed at a similar low spectral resolution 
(filled blue).  The rest equivalent width of \GBsixteen\ is shown as a blue narrow-hatched histogram 
\citep[taken from][]{schirmer2016}.  
The restframe equivalent widths of low-z \lya\ nebulae 
are significantly higher than those of Type 2 AGN, but comparable to those of Green Peas. 
}
\label{fig:type2rew}
\end{figure}

The low-z \lya\ nebulae were first selected, in part, due to their unusual location in color-color space.
In Figure~\ref{fig:type2colors}, we plot the measured $g-r$ versus $r-i$ colors of the 
low-z \lya\ nebula sample along with the colors for Type 2 AGN at similar redshifts.  
Broadband colors are affected by strong emission lines, e.g., 
\oii, \oiii, \halpha, falling into different bands depending on the redshift.  
In particular, at $z\approx0.3$, 
the \oiii\ doublet -- the strongest optical emission lines for these sources -- is contained within 
the $r$ bandpass, leading to blue $r-i$ and red $g-r$ colors, and a greenish appearance in 
SDSS $gri$ color composite images.  The low-z \lya\ nebulae were selected from 
this portion of the color-color diagram.  Type 2 AGN at 
similar redshifts show less extreme colors, 
suggesting they may have lower \oiii\ equivalent widths.  
Figure~\ref{fig:type2rew} 
confirms this, showing that the restframe \oiii\ equivalent widths 
for the low-z \lya\ nebula pilot sample are 
statistically higher than those of typical Type 2 AGN 
(KS test p-value of $4\times10^{-5}$), 
but comparable to those of Green Peas (KS test p-value of 0.12).  
The weaker \halpha\ line also has an effect on the broadband colors, 
as it shifts from the $i$-band to the $z$-band at $z\sim0.296$, leading 
to somewhat bluer $r-i$ and redder $r-z$ colors for the higher redshift 
end of the samples \citep{schirmer2016}.
 
\begin{figure}
      \begin{adjustbox}{addcode={\begin{minipage}{\width}}{\caption{
Stacked low resolution spectrum of the low-z \lya\ nebulae 
in our pilot sample (dark blue) vs. the stacked spectra of {\bf star-forming} Green Peas 
(green) and Type 2 AGN (magenta), normalized to the region around 5995-6005\AA.
The low-z \lya\ nebula 
stack shows substantially bluer restframe optical continuum 
than that of the Type 2 AGN, 
although not as blue as that of Green Peas.  (Inset) Zoom-in on the 
region around 4000\AA.  
While the Ca H absorption line is affected by emission filling from 
the \neiii$\lambda$3968\AA\ emission line at nearly the same wavelength, 
the Ca K line is clearly weaker in the low-z \lya\ nebulae stack than in 
the Type 2 AGN stack, consistent with low-z \lya\ nebulae hosting 
younger stellar populations on average.
\label{fig:stackspec}
      	}\end{minipage}},rotate=90,center}
	\includegraphics[angle=0,width=8in]{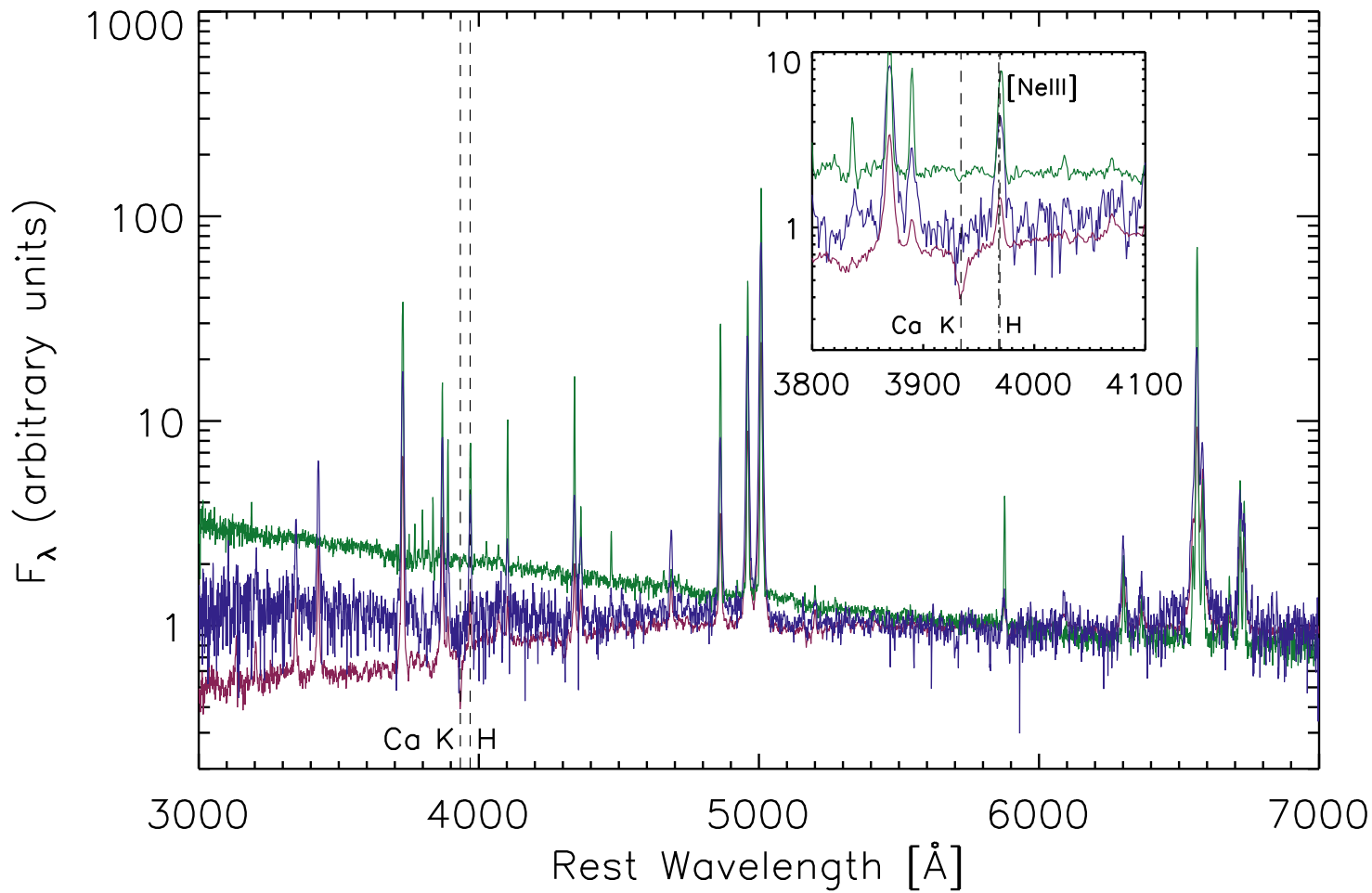}
	\end{adjustbox}
\end{figure}

\begin{figure}
\includegraphics[angle=0,width=4.5in]{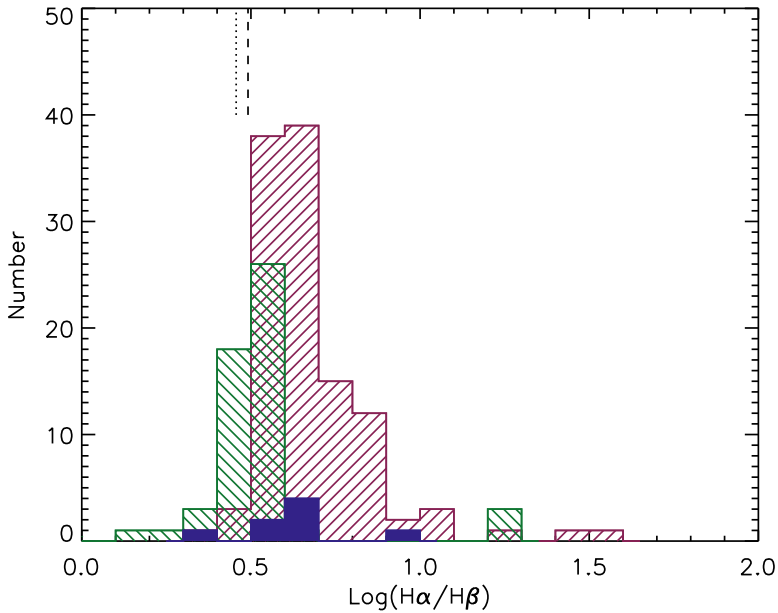}
\caption[]{Balmer decrements measured using low resolution spectra of the pilot low-z \lya\ nebulae 
sample (dark blue filled histogram) as compared 
to Type 2 AGN \citep[magenta hatched; taken from][]{bolton2012} 
and Green Peas \citep[green hatched; taken from][]{thomas2013}.
The expected values for Case B conditions, assuming no dust extinction, of $\sim$3.1 (AGN; dashed line) 
and $\sim$2.86 (star formation; dotted line) are indicated \citep{dopita2003}.  
Low-z \lya\ nebulae 
have Balmer decrements comparable to Type 2 AGN, suggesting that dust extinction is not the 
primary explanation for the observed difference in optical continuum color between the two populations.
}
\label{fig:histhahb}
\end{figure}

However, it is not emission lines alone that cause 
low-z \lya\ nebulae to have unusual colors; the optical continuum also plays a role.  
To explore this, we next stack the low resolution spectra 
for the low-z \lya\ nebula pilot sample (8 objects, 
including the SDSS spectrum of \GBfour), and 
SDSS/BOSS spectra for the Green Peas (71 objects) 
and Type 2 AGN (117 objects) comparison samples. 
After redshifting to the restframe and normalizing the spectra at 5995-6005\AA, 
a region free of strong lines, we generate a stacked spectrum 
using a geometric mean, as this has been 
found to preserve the shape of the continuum better than a simple median \citep{vandenberk2001}.  
Using a geometric mean does require masking negative pixels during the stacking; 
however, this is a small effect, with 
fewer than 1-2\% of the pixels being masked in this procedure.  
In addition, we reach the same basic conclusions if we instead use a simple median stack. 
The results of the stacking are shown in Figure~\ref{fig:stackspec}, where 
the low-z \lya\ nebula stacked spectrum shows a much bluer restframe 
optical continuum than the stacked Type 2 AGN.  
This difference could in principle come from dust extinction, from 
scattered light from the central engine, or from the underlying stellar continuum of the host galaxy.  
As a check on the role of dust, in Figure~\ref{fig:histhahb} we 
plot the observed \halpha/\hbeta\ ratio (i.e., the Balmer decrement) for the low-z \lya\ nebula 
pilot sample, the Type 2 AGN, and the Green Peas.  
All three populations show comparable Balmer decrements; 
although there is a hint that the low-z \lya\ nebulae 
may show slightly lower \halpha/\hbeta\ ratios compared to Type 2 AGN 
-- consistent with lower dust extinction -- this result is not statistically 
significant given the current sample size (KS test p-value of 0.51).  
Excess blue continuum has been found previously in higher luminosity 
Type 2 AGN samples \citep[e.g.,][]{zakamska2003}, with the authors ruling out 
a strong contribution from scattered light based on the lack of broad components on 
the permitted line emission.  In Section~\ref{sec:powering}, we showed 
that the low-z \lya\ nebula sample also lack broad permitted line emission, arguing 
against the AGN scattered light explanation. Furthermore, \citet{schirmer2016} argue, 
based on SED fitting for one low-z \lya\ nebula (\GBsixteen) and on a 
color analysis of neighboring galaxies for three more systems, that the continuum in the ultraviolet, 
at even bluer wavelengths, is fully explained by a combination 
of stellar light and nebular continuum, leaving no room for a scattered AGN component.   
Thus, it appears likely that much of the observed color difference is due 
instead to differences in the underlying stellar continuum of the host galaxy 
between low-z \lya\ nebulae and typical Type 2 AGN.  This is corroborated by the much weaker 
Ca K stellar absorption feature in the low-z \lya\ nebula stacked spectrum 
as compared to the Type 2 AGN stack (Figure~\ref{fig:stackspec}). 

\subsection{Sizes and Concentrations}
\label{sizeconcen}

\begin{figure}
\includegraphics[angle=0,width=4.5in]{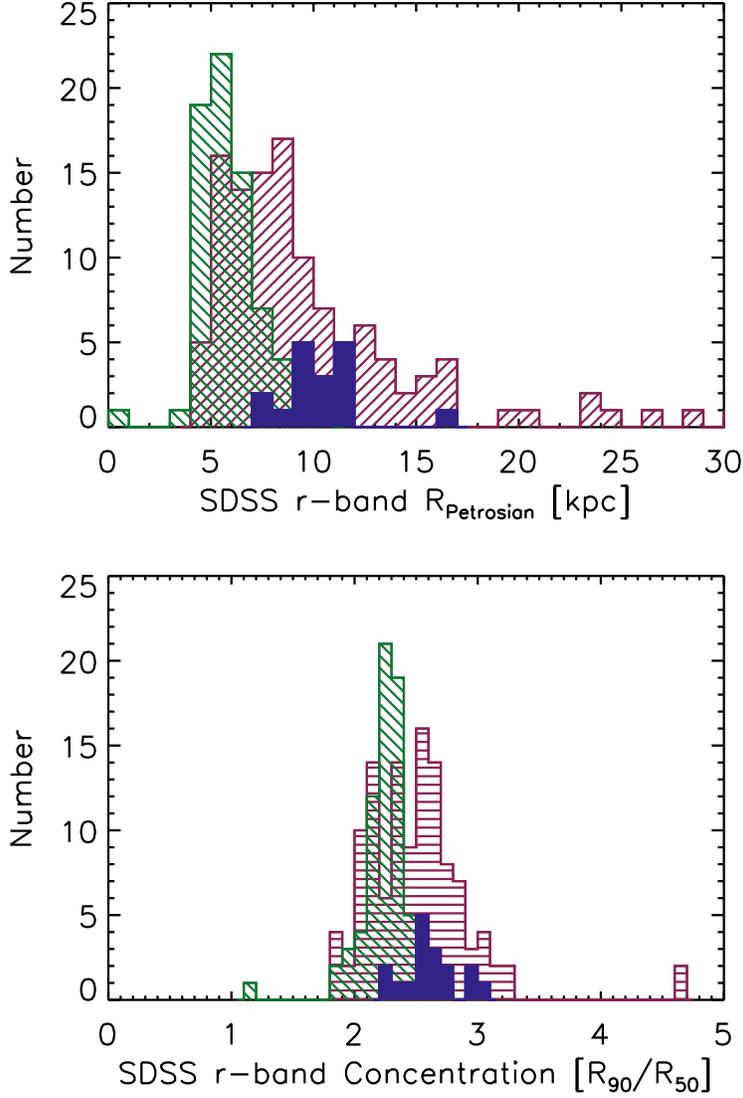}
\caption[]{
Histograms of sizes and concentrations for Grean Peas (hatched green), Type 2 AGN (hatched magenta), 
and low-z \lya\ nebulae (filled blue), as measured from SDSS $r$-band imaging.  
(Top) SDSS $r$-band Petrosian radii. (Bottom) Concentration estimated using the ratio between the radii 
containing 90\% and 50\% of the SDSS $r$-band Petrosian flux ($R_{90}/R_{50}$).
The SDSS $r$-band contains the bright \oiii\ emission line at these redshifts, thus the $r$-band estimates are 
a good proxy for the relative nebular sizes and concentrations of these three populations.  
Low-z \lya\ nebulae 
appear larger and less concentrated than Green Pea galaxies, 
as expected, but they are comparable in size and concentration to Type 2 AGN.  
}
\label{fig:sizeconcen}
\end{figure}

The low-z \lya\ nebulae were selected not simply due to their unusual location in color-color space, 
but also due to larger sizes in the SDSS $r$-band imaging.  
As broadband sizes naturally underestimate the true nebular size, deep spectroscopy or 
narrowband imaging is required to assess the full extent of the \oiii\ emission.
However, since at $z\approx0.3$ the high equivalent width \oiii\ doublet is contained 
within the $r$-band filter (contributing $\sim$20-60\% of the $r$-band flux 
for the pilot low-z \lya\ nebula sample), the $r$-band imaging can give us a sense for the 
relative sizes and concentrations of low-z \lya\ nebulae versus other extreme emission line samples.  
As a proxy for the size, we therefore use the $r$-band Petrosian radius, and for 
concentration, we use the ratio of the $r$-band 90\% and 50\% Petrosian radii 
derived from the SDSS database.  
Figure~\ref{fig:sizeconcen} shows the distribution of measured $r$-band sizes and concentrations 
for low-z \lya\ nebulae, Green Peas, and Type 2 AGN at similar redshifts.  
By construction, Green Peas are small and compact, particularly when compared to the low-z 
\lya\ nebulae.  
Type 2 AGN, by comparison, are less compact and show a much wider range of sizes, 
overlapping both of the other two populations.  
Low-z \lya\ nebulae are similar in concentration to Type 2 AGN, 
and are comparable to the larger size tail of 
the Type 2 AGN distribution.  Thus, while low-z \lya\ nebulae 
are indeed noticeably spatially extended 
in the $r$-band imaging (the data originally used to isolate the low-z \lya\ nebula 
sample), they do not appear to be the largest such objects at these redshifts.  

\begin{figure}
\includegraphics[angle=0,width=4.5in]{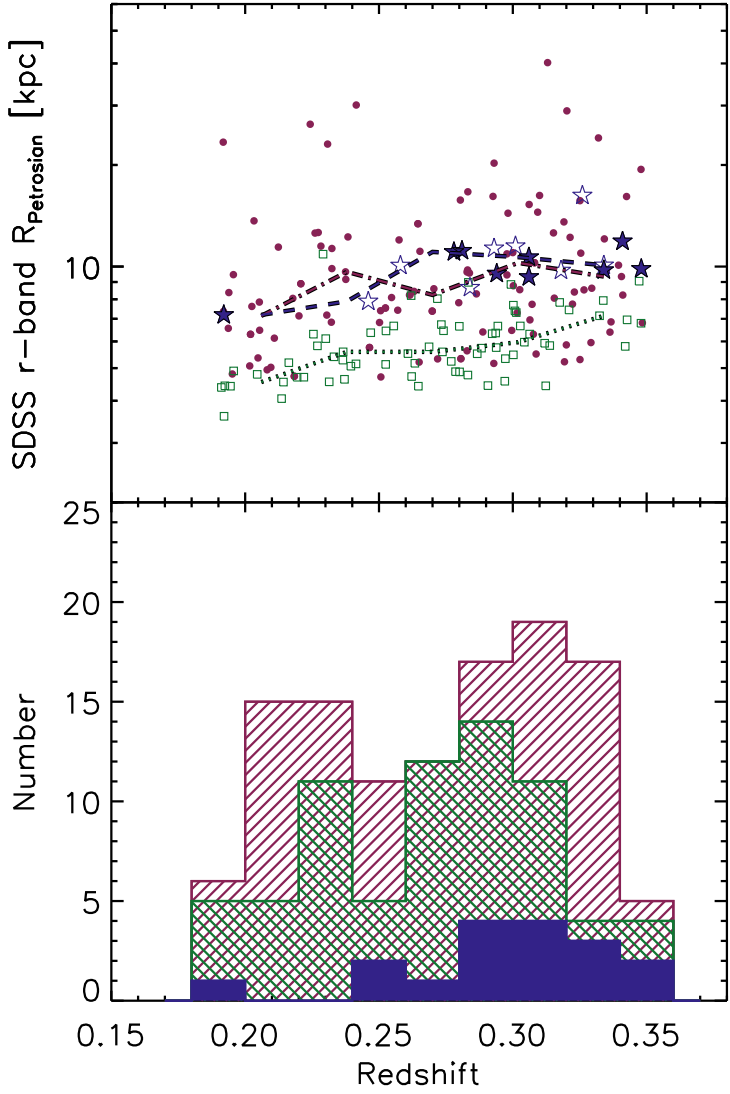}
\caption[]{
(Top) Redshift versus physical $r$-band Petrosian radius 
for Grean Peas (green squares), Type 2 AGN (magenta circles), 
and low-z \lya\ nebulae 
(blue stars), as measured from SDSS $r$-band imaging.  
The low-z \lya\ nebulae 
targeted in this paper are indicated as filled blue stars.  
The median radius as a function of redshift is shown with colored lines 
for the Type 2 AGN (magenta dot-dashed), Green Peas (green dotted), 
and low-z \lya\ nebula (blue dashed) samples.  
(Bottom) Redshift distribution for the Green Peas (hatched green histogram), 
Type 2 AGN (hatched magenta), and the low-z \lya\ nebulae (filled blue).
Although the low-z \lya\ nebula sample is skewed towards higher redshifts 
than the Type 2 AGN sample, the median size does not vary significantly with redshift 
and the largest objects are drawn from a range of redshifts.
}
\label{fig:zVSpetroradkpczdist}
\end{figure}

The presence of a tail of Type 2 AGN with large $r$-band sizes suggests that there may 
be an additional sample of low-z \lya\ nebulae just outside the original 
``Green Bean'' selection window. 
However, this large size tail could simply be due to lower redshift sources 
for which more extended emission is visible.  
To check for biases in the redshift distribution of the different samples, 
in Figure~\ref{fig:zVSpetroradkpczdist} we plot the redshift versus $r$-band Petrosian radii for all sources. 
The low-z \lya\ nebulae are indeed skewed towards higher redshifts, but 
the median size for both populations does not vary significantly with redshift, and 
the large size tail of the Type 2 AGN distribution is populated by objects 
from a range of redshifts.

\begin{figure}
\includegraphics[angle=0,width=4.5in]{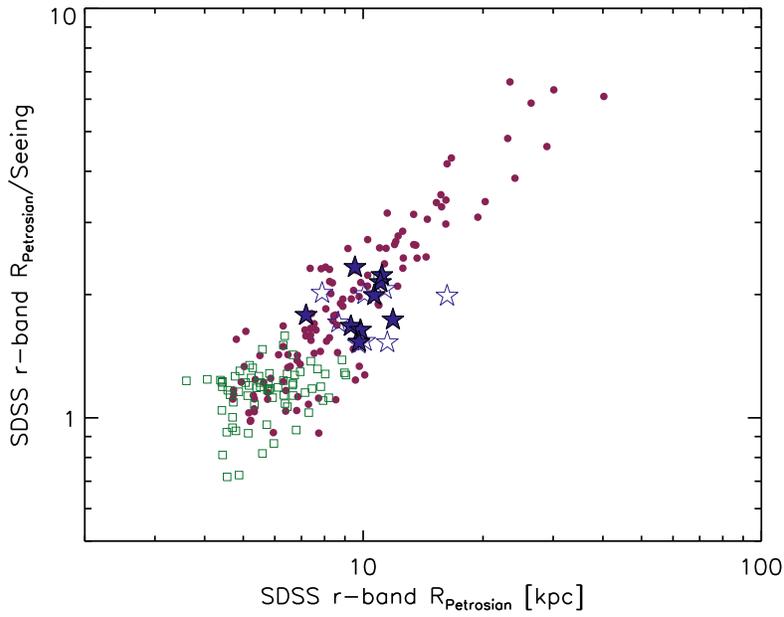}
\caption[]{
Ratio of the observed $r$-band Petrosian radius (in arcseconds) to the $r$-band seeing 
versus the physical $r$-band Petrosian radius (in kiloparsecs) for 
Grean Peas (green squares), Type 2 AGN (magenta circles), 
and low-z \lya\ nebulae 
(blue stars), as measured from SDSS $r$-band imaging.  
The low-z \lya\ nebulae 
targeted in this paper are indicated as filled blue stars.  
The tail to large sizes is populated by sources that are well beyond 
the size of the seeing.
}
\label{fig:petroradkpcVSseeing}
\end{figure}

It is also possible that seeing effects could bias the size measurements.  The low 
redshift \lya\ nebula sample was explicitly cleaned of poor seeing 
targets \citep{schirmer2016}; to check for poor seeing in the other samples, 
we look at the observed $r$-band Petrosian radius (in arcseconds) scaled by 
the size of the seeing versus the physical $r$-band Petrosian radius 
(Figure~\ref{fig:petroradkpcVSseeing}).  
As expected, the Green Peas are comparable in size to the seeing, while the Type 2 AGN 
and low-z \lya\ nebula samples extend to much larger physical sizes and 
to larger ratios of physical size to seeing.  The seeing 
does not appear to be important for these larger sources, as there are no sources 
in the lower right quadrant of the plot, i.e.,  with large measured physical sizes but 
with low ratios of physical size to seeing.

Another potential bias is the fact that, as we showed in Section~\ref{sec:colors}, 
the Type 2 AGN have slightly lower 
\oiii\ equivalent widths on average than the low-z \lya\ nebulae.  
Since \oiii\ therefore contributes a smaller fraction of the $r$-band flux, 
the $r$-band size measurement will be less sensitive to extended line emission 
for these sources.  However, this does not explain the large size tail of the 
Type 2 AGN sample as it would if anything tend to bias the $r$-band size measurement towards 
lower sizes.

In carrying out this analysis, we made use of SDSS Petrosian radii 
in order to minimize redshift biases \citep{petrosian1976,blanton2001,yasuda2001}.  
However, in cases with a particularly compact central core embedded within a more 
extended nebula -- as may be the case in the Type 2 AGN sample -- the Petrosian radius 
measurement will be dominated by the core profile.
Again, however, this effect will tend to bias the $r$-band size measurements to smaller values.  
Thus, we conclude that the large measured sizes are likely to be physical and warrant further study.  
We will present an analysis of this population of extended Type 2 AGN sources in a 
subsequent paper.

\subsection{Kinematics}
\label{sec:kinematics}

\begin{figure}
\includegraphics[angle=0,width=6in]{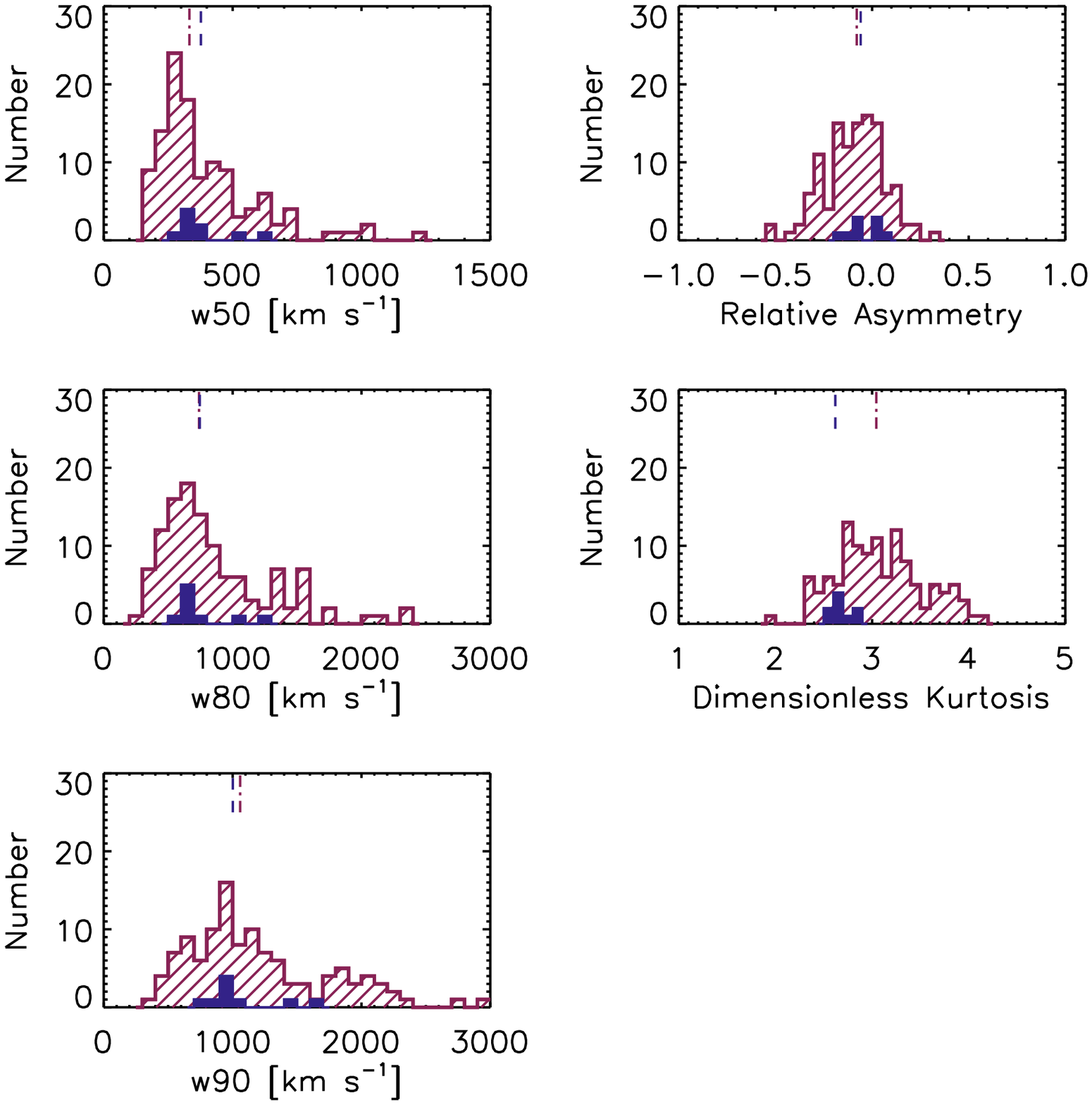}
\caption[]{
Histogram of $w50$ (top), $w80$ (middle), and $w90$ (bottom) for the low-z \lya\ nebulae 
(medium resolution data; filled blue histogram), with the \citet{yuan2016} 
Type 2 AGN comparison sample shown (hatched magenta).  
The short colored line along the top axis indicates the median value for the low-z \lya\ nebula 
(blue dashed) and Type 2 (magenta dot-dashed) samples, respectively.  The 
low-z \lya\ nebulae do not stand out relative to typical Type 2 AGN 
in terms of \oiii\ kinematics. 
}
\label{fig:oiiikin1}
\end{figure}

To compare the kinematics of our low-z \lya\ nebula 
sample with that of typical Type 2 AGN, 
we use the non-parametric kinematic measurements from Section~\ref{sec:fitting}, 
derived using multiple Gaussian component fitting of the \oiii\ doublet line profiles in the 
same manner as \citet{yuan2016}, after smoothing our medium resolution data to 
the same spectral resolution as the Type 2 AGN sample data.  
Figure~\ref{fig:oiiikin1} shows histograms of all measured parameters for both 
the Type 2 AGN and the pilot low-z \lya\ nebula sample.  
In terms of the measured velocity widths -- $w50$, $w80$, and $w90$ -- 
a two-sided KS test reveals no difference between the two samples (p-values of 0.28, 0.57, and 0.28).  
The relative asymmetry ($R$) distribution for the low-z \lya\ nebulae 
also shows no significant difference 
as compared to the typical Type 2 AGN sample (KS test p-value of 0.45).  
The dimensionless kurtosis (r9050) measurements 
do appear to diverge slightly from those of the Type 2 sample (KS test p-value of 0.01), 
in the sense that low-z \lya\ nebulae show lower kurtosis (i.e., smaller tails) on average than typical Type 2 AGN.  
The effect is not strong, but to the extent that a high dimensionless kurtosis signals 
the presence of faster outflows (i.e., more prominent tails on the \oiii\ line profile), 
this may suggest that, at the very least, low-z \lya\ nebulae 
are not dominated by above-average outflow speeds compared to other Type 2 AGN.

\section{Discussion}
\label{sec:discussion}

Low-z \lya\ nebulae appear to host and be powered by Type 2 AGN, with consistent narrow-line 
kinematics, emission line flux ratios, and ionizing spectral slopes.  
They show more extreme colors than either Green Peas or Type 2 AGN, 
partly due to strong \oiii\ doublet emission within the $r$-band, but also 
due to differences in optical continuum colors.  
Low-z \lya\ nebulae are larger and less compact than Green Peas (by construction), 
and they are comparable to the overall Type 2 AGN population in terms of 
$r$-band concentration and size, although skewed towards the larger 
size end of the Type 2 AGN size distribution.  
They show higher than average 
mid-infrared luminosity, high \oiii\ rest equivalent widths, and bluer 
optical continua, but with outflow kinematics that are 
typical for Type 2 AGN. 

These findings may imply that rather than being a distinct population, 
these low-z \lya\ nebulae 
are drawn from the high luminosity end of the Type 2 AGN population.  
Previous studies have shown that the \oiii\ restframe equivalent width is correlated 
with the overall \oiii\ luminosity for Type 2 AGN \citep{zakamska2003}.  
The presence of blue optical continua in Type 2 AGN (i.e., bluer than that of an old stellar population) 
has received a fair amount of attention in the literature \citep[e.g.,][]{gonzalez1993,heckman1995,cidfernandes2001,kauffmann2003,zakamska2003}, 
and is thought to be a result of a natural correlation between nuclear activity and 
starburst activity in the host galaxy, such that 
more luminous Type 2 AGN are accompanied by an overall younger stellar population.  
In addition, there are indications that the optical continuum color gets bluer with 
increasing \oiii\ luminosity \citep{zakamska2003}.  
Since low-z \lya\ nebulae 
appear to be drawn from the high luminosity end of the Type 2 AGN population, 
we would expect them to exhibit bluer optical continuum and higher \oiii\ equivalent widths.  
It is plausible therefore that the low-z \lya\ nebulae 
phenomenon is related to high nuclear activity, 
which is correlated with high mid-infrared luminosity, more star formation, 
bluer optical continua, and higher \oiii\ equivalent widths, 
all of which leads to an object being selected as a ``Green Bean'' 
in color-color space.

Despite being some of the more luminous Type 2 AGN at this redshift, the 
low-z \lya\ nebulae 
are not dramatically larger than other Type 2 AGN, at least in terms of broadband sizes.  
If the broadband measurements are giving us a correct impression of the relative sizes, 
this is consistent with findings that the size of the AGN narrow-line region (NLR) scales 
with the power of the central engine only out to a characteristic radius, 
after which point, the size-luminosity relation seems to flatten 
\citep[e.g.,][]{liu2013a,hainline2013,hainline2014,sun2017}.  
Early explanations for this flattening suggested that the maximum size of the NLR 
is set, not by the available energy from the central engine, but rather by the extent 
of the gas that is at the right density and ionization state to produce \oiii\ emission.  
Beyond this characteristic radius, the gas is too low density or too highly ionized to produce 
observable \oiii\ -- effectively a matter-bounded scenario.  More recent modeling has suggested 
that this matter-bounded effect alone may not be sufficient to explain the turn-over in 
the size-luminosity 
relation, instead indicating that it may be due to the effect of optically thick clouds 
shielding gas beyond a certain distance from the ionizing impact of the AGN \citep{dempsey2018}. 
Putting low-z \lya\ nebulae 
in context with this previous work on the size-luminosity relation will 
require measuring the nebular sizes of this sample 
using a consistently applied size definition \citep[e.g.,][]{liu2013a}. 
None-the-less, the relative sizes derived from SDSS broadband imaging are consistent with 
the picture in which the luminosity is not the sole driver of the extent of the NLR, even 
for luminous Type 2 AGN like those powering low-z \lya\ nebulae.  

This work motivates further imaging and spectroscopic follow-up of both low-z \lya\ nebulae 
and other luminous Type 2 AGN.
Previous work has suggested that the central engine in low-z \lya\ nebulae 
may be in the process of shutting down, 
leaving behind extended emission line ionization echoes \citep{schirmer2013,schirmer2016,kawamuro2017}.  
If true, studying low-z \lya\ nebulae 
and other Type 2 AGN at similar luminosities may provide insight into the mechanisms and conditions 
responsible for triggering such a ramp down in nuclear activity. 

In being AGN powered, low-z \lya\ nebulae 
are similar to high redshift \lya\ nebulae, which often show evidence 
of AGN activity \citep[e.g.,][]{overzier13,pres15b}.
Yet, there are also hints that low-z \lya\ nebulae 
show some distinct differences.  
\citet{schirmer2016} found that the environments of low-z \lya\ nebulae 
were often average or even underdense, 
in contrast to the generic finding that high redshift \lya\ nebulae reside in overdense regions 
that are in the process of forming galaxy groups and clusters 
\citep[e.g.,][]{sai06,pres08,yang09,yang10}.  While high redshift \lya\ nebulae 
that have been studied at high spatial resolution appear to be messy, asymmetric, 
dynamically unrelaxed regions containing both 
high angular momentum gas and many disky, star-forming galaxies \citep[e.g.,][]{pres12b,pres15a}, 
low-z \lya\ nebulae are for the most part centered on a single galaxy, often with signs of bipolar 
outflows \citep{davies2015,schirmer2016,kawamuro2017}.  
More extensive follow-up of the emission line luminosities, kinematics, 
and environments of extended nebulae at both low and high redshift will be important to for better understanding 
the relationship between the two populations and, potentially, for providing new constraints on the evolution 
of the circumgalactic medium and galaxy formation physics across cosmic time.


\section{Conclusions}
\label{sec:conclusions}

We present follow-up spectroscopy for a subset of the $z\approx0.3$ \lya\ nebulae from \citet{schirmer2013,schirmer2016}.  
Our spectroscopic results show that low-z \lya\ nebulae 
are powered by Type 2 AGN, with high excitation line 
emission, relatively narrow linewidths (${\rm FWHM}\lesssim1000$ km s$^{-1}$), and typical ionizing continua.  
They do appear to have intrinsically bluer optical continua and higher \oiii\ equivalent widths than 
Type 2 AGN at the same redshift, both of which are consistent with low-z \lya\ nebulae 
being more luminous overall than the average Type 2 AGN at the same redshift.  
Low-z \lya\ nebulae have larger $r$-band sizes and lower $r$ concentrations than star-forming Green Pea galaxies, 
but are comparable in $r$-band size and concentration to the larger Type 2 AGN population.  
Using the \oiii\ emission line profiles, 
we compared the kinematics of the narrow-line region gas in these objects 
to Type 2 AGN, finding that low-z \lya\ nebulae 
kinematically resemble typical Type 2 AGN in terms of 
a variety of non-parametric measures.  The only kinematic difference between the low-z \lya\ nebula 
and Type 2 samples seems to be in the dimensionless kurtosis measurements, in 
the sense that the low-z \lya\ nebula \oiii\ emission line profiles have slightly weaker tails.  
Based on the existing pilot study, it appears that low-z \lya\ nebulae 
represent a high luminosity subset of normal 
Type 2 AGN.  More extensive spectroscopic follow-up of the full low-z \lya\ nebulae 
sample will be needed to further investigate these issues and to understand how they relate to \lya\ 
nebulae at higher redshifts. 

\acknowledgments
The authors would like to thank Kristian Finlator and Steve Finkelstein for helpful discussions, 
Christy Tremonti, Eric Wilcots, and the University of Wisconsin-Madison Astronomy Department for 
their kind hospitality while we were working on this project, and the anonymous referee for 
suggestions that improved the paper. 

Funding for SDSS-III has been provided by the Alfred P. Sloan Foundation, the Participating Institutions, the National Science Foundation, and the U.S. Department of Energy Office of Science. The SDSS-III web site is http://www.sdss3.org/.  SDSS-III is managed by the Astrophysical Research Consortium for the Participating Institutions of the SDSS-III Collaboration including the University of Arizona, the Brazilian Participation Group, Brookhaven National Laboratory, Carnegie Mellon University, University of Florida, the French Participation Group, the German Participation Group, Harvard University, the Instituto de Astrofisica de Canarias, the Michigan State/Notre Dame/JINA Participation Group, Johns Hopkins University, Lawrence Berkeley National Laboratory, Max Planck Institute for Astrophysics, Max Planck Institute for Extraterrestrial Physics, New Mexico State University, New York University, Ohio State University, Pennsylvania State University, University of Portsmouth, Princeton University, the Spanish Participation Group, University of Tokyo, University of Utah, Vanderbilt University, University of Virginia, University of Washington, and Yale University.


\begin{deluxetable}{ccccccccc}
\tabletypesize{\scriptsize}
\rotate
\tablecaption{Observing Log}
\tablewidth{0pt}
\tablehead{
\colhead{UT Date} & \colhead{SDSS Name} & \colhead{Position} & \colhead{Grating} & \colhead{Spectral} & \colhead{Spatial} & \colhead{Exposure} & \colhead{Seeing} & \colhead{Conditions} \\
 &  & Angle& Wavelength &Resolution\tablenotemark{a} & Bin (pix) & Time (sec) & (arcsec) & }
\startdata
\\
2018 Jan 12 & SDSSJ011341.11+010608.5 &   129.00$^{\circ}$ & 4800\AA\ / 6600\AA\ & 1.7\AA\ /   1.8\AA\ &  1 &     5x600 &    1.0-1.6 &                          Clear\\
  &  &  85.00$^{\circ}$ &  & &  &      3x600 &  & \\
 & SDSSJ015930.84+270302.2 &   155.00$^{\circ}$ &  & &  &      6x600 &  & \\
  &  &  28.50$^{\circ}$ &  & &  &      4x600 &  & \\
\\
\hline\\
2018 Jan 24 & SDSSJ115544.59$-$014739.9 &  25.93$^{\circ}$ &  4800\AA\ / 6600\AA\ & 1.7\AA\ /   1.9\AA\ &  1 &      4x600 &    1.5-1.9 &                          Clear\\
  &  &   132.56$^{\circ}$ &  &  &   &   4x600 &  & \\
 & SDSSJ134709.12+545310.9 &   138.99$^{\circ}$ &  &  &   &   4x600 &  & \\
  &  &  60.35$^{\circ}$ &  &  &    &  4x600 &  & \\
 & SDSSJ150517.63+194444.8 &   88.40$^{\circ}$ &  &  &   &   2x600 &  & \\
\\
\hline\\
2018 Apr 15 & SDSSJ150517.63+194444.8 &   173.54$^{\circ}$ & 4800\AA\ / 6600\AA\ &  1.7\AA\ /   1.5\AA\ &  2 &      2x600 &        1.0 &                          Clear\\
 & SDSSJ135155.48+081608.4 &   152.27$^{\circ}$ &  &  &    &  4x600 &  & \\
  &  &  46.11$^{\circ}$ &  &  &    &  4x600 &  & \\
 & SDSSJ145533.69+044643.2 &   102.51$^{\circ}$ &  &  &    &  3x600 &  & \\
  &  &  30.56$^{\circ}$ &  &  &    &  3x600 &  & \\
\\
\hline\\
2018 Apr 22 & SDSSJ144110.95+251700.1 &  2.79$^{\circ}$ & 4562\AA\ / 6137\AA\ & 1.7\AA\ /   1.5\AA\ &  1 &     3x600 &        1.3 &                         Cloudy\\
 & SDSSJ150420.68+343958.2 &   108.46$^{\circ}$ &  &  & &  6x300 &  & \\
\\
\hline\\
2018 May 09 & SDSSJ115544.59$-$014739.9 &   171.28$^{\circ}$ & 4400\AA\ / 7500\AA\ & 5.9\AA\ /   6.5\AA\ &  1 &   4x200 &    1.2-1.3 &                          Clear\\
 & SDSSJ134709.12+545310.9 &   138.99$^{\circ}$ &  &  &   &   3x200 &  & \\
 & SDSSJ135155.48+081608.4 &   152.27$^{\circ}$ &  &  &   &   3x200 &  & \\
 & SDSSJ144110.95+251700.1 &  2.79$^{\circ}$ &  &  &   &  3x300 &  & \\
 & SDSSJ145533.69+044643.2 &   102.51$^{\circ}$ &  &  &   &   3x200 &  & \\
 & SDSSJ150420.68+343958.2 &   108.46$^{\circ}$ &  &  &   &  4x300 &  & \\
 & SDSSJ150517.63+194444.8 &   88.40$^{\circ}$ &  &  &   &   3x200 &  & \\
\enddata
\tablecomments{All observations taken using a 1.5\arcsec\ slit.}
\tablenotetext{a}{Values listed for blue / red gratings at nominal wavelengths of 4800\AA\ / 6500\AA.}
\label{tab:obs}
\end{deluxetable}


\end{document}